\documentclass[
]{ceurart}

\usepackage{listings}
\usepackage{colortbl}
\usepackage[flushleft]{threeparttable} 
\usepackage{booktabs, caption}
\usepackage{graphicx}
\usepackage{adjustbox}

\begin{document}

\copyrightyear{2023}
\copyrightclause{Copyright for this paper by its authors.
  Use permitted under Creative Commons License Attribution 4.0
  International (CC BY 4.0).}

\conference{ReNeuIR'23: Workshop on Reaching Efficiency in Neural Information Retrieval,
  July 23--27, 2023, Taipei, Taiwan}

\title{Injecting Domain Adaptation with Learning-to-hash for Effective and Efficient Zero-shot Dense Retrieval}

\author[1]{Nandan Thakur}[]
\address[1]{David R. Cheriton School of Computer Science, University of Waterloo}

\author[2]{Nils Reimers}[]
\address[2]{Cohere.ai}

\author[1]{Jimmy Lin}[]


\begin{abstract}
Dense retrieval overcome the lexical gap and has shown great success in ad-hoc information retrieval (IR). 
Despite their success, dense retrievers are expensive to serve across practical use cases.
For use cases requiring to search from millions of documents, the dense index becomes bulky and requires high memory usage for storing the index. 
More recently, learning-to-hash (LTH) techniques, for e.g., BPR and JPQ, produce binary document vectors, thereby reducing the memory requirement to efficiently store the dense index. 
LTH techniques are supervised and finetune the retriever using a ranking loss.
They outperform their counterparts, i.e., traditional out-of-the-box vector compression techniques such as PCA or PQ.
A missing piece from prior work is that existing techniques have been evaluated only in-domain, i.e., on a single dataset such as MS MARCO. 
In our work, we evaluate LTH and vector compression techniques for improving the downstream zero-shot retrieval accuracy of the TAS-B dense retriever while maintaining efficiency at inference. 
Our results demonstrate that, unlike prior work, LTH strategies when applied naively can underperform the zero-shot TAS-B dense retriever on average by up to 14\% nDCG@10 on the BEIR benchmark.
To solve this limitation, in our work, we propose an easy yet effective solution of injecting domain adaptation with existing supervised LTH techniques. 
We experiment with two well-known unsupervised domain adaptation techniques: GenQ and GPL. Our domain adaptation injection technique can improve the downstream zero-shot retrieval effectiveness for both BPR and JPQ variants of the TAS-B model by on average 11.5\% and 8.2\% nDCG@10 while both maintaining $32\times$ memory efficiency and $14\times$ and $2\times$ speedup respectively in CPU retrieval latency on BEIR.
All our code, models, and data are publicly available at \url{https://github.com/thakur-nandan/income}.
\end{abstract}


\maketitle

\vspace{-3mm}
\section{Introduction}
Dense retrieval has become a core component within several downstream NLP and web-search tasks such as question-answering \cite{lee-etal-2019-latent, karpukhin-etal-2020-dense}, semantic similarity \cite{reimers-gurevych-2019-sentence}, conversational search \cite{10.1145/3404835.3462856}, entity retrieval \cite{gillick-etal-2019-learning}, fact-checking \cite{samarinas-etal-2021-improving} and passage retrieval \cite{complimenting2020, xiong2020approximate, Hofstaetter2021_tasb_dense_retrieval, 10.1145/3404835.3462880, 10.1162/tacl_a_00369}. 
In dense retrieval, dual-encoder models encode semantically correlated queries and documents and represent them as spatially close embeddings, i.e., dense vector representations.
Dense retrieval can be efficiently conducted via approximate nearest neighbor (ANN) search at inference \cite{johnson2019billion}. 

Dual-encoder models encode and store document embeddings within an index, which starts to become bulky, once the documents start to increase within the corpus. 
It becomes expensive to serve dense retrievers practically. 
For example, storing 21M (million) passages within a dense index requires about 65~GB of memory \cite{yamada2021efficient}. 
As nearest-neighbor (exact) search in large vector spaces is rather slow, ANN techniques such as HNSW \cite{MALKOV201461} are popularly used, which results in even higher memory requirements such as 150 GB \cite{yamada2021efficient}. 
As a result, for searching on a very large corpus, we require a high-memory and expensive machine to host the bulky index, which hinders the practical application of dense retrieval.

Aware of the issue, there have been recent efforts to improve the efficiency of dense retrieval models by compressing the vector representation to lower memory, i.e., space requirements. 
Traditionally, out-of-the-box unsupervised strategies such as dimension reduction with principal component analysis (PCA) or product quantization (PQ) \cite{jegou2010product} have been popular for retrieval \cite{izacard2020memory, ma-etal-2021-simple}. More recently, supervised learning-to-hash (LTH) techniques have been introduced such as BPR \cite{yamada2021efficient} and JPQ \cite{zhan2021jointly}. LTH techniques map the original dense embeddings into a Hamming space (binary space) and obtain the low-dimensional binary codes. Unfortunately, all existing vector compression methods are proposed under the single-domain retrieval assumption. However, in practice, many retrieval applications span across various domains with diverse distributions; often with scarce training data \cite{thakur2021beir}. 

In our work, we focus on the Pareto frontier between downstream zero-shot retrieval accuracy while maintaining the efficiency of compression methods. We evaluate TAS-B \cite{Hofstaetter2021_tasb_dense_retrieval}, a robust generalizable dense retriever on 18 diverse retrieval datasets on the BEIR benchmark \cite{thakur2021beir}. 
We evaluate out-of-the-box vector compression techniques to efficiently compress the embeddings, with the least drop in zero-shot retrieval effectiveness. 
We show that prior work on hashing strategies focused on single-domain retrieval fails severely under domain shifts in zero-shot retrieval, and can underperform the zero-shot TAS-B dense retriever on average by up to 14\% nDCG@10 on average on BEIR.

\begin{figure*}[t]
\centering
\begin{center}
    \includegraphics[trim=0 0 0 0,clip,width=0.9\textwidth]{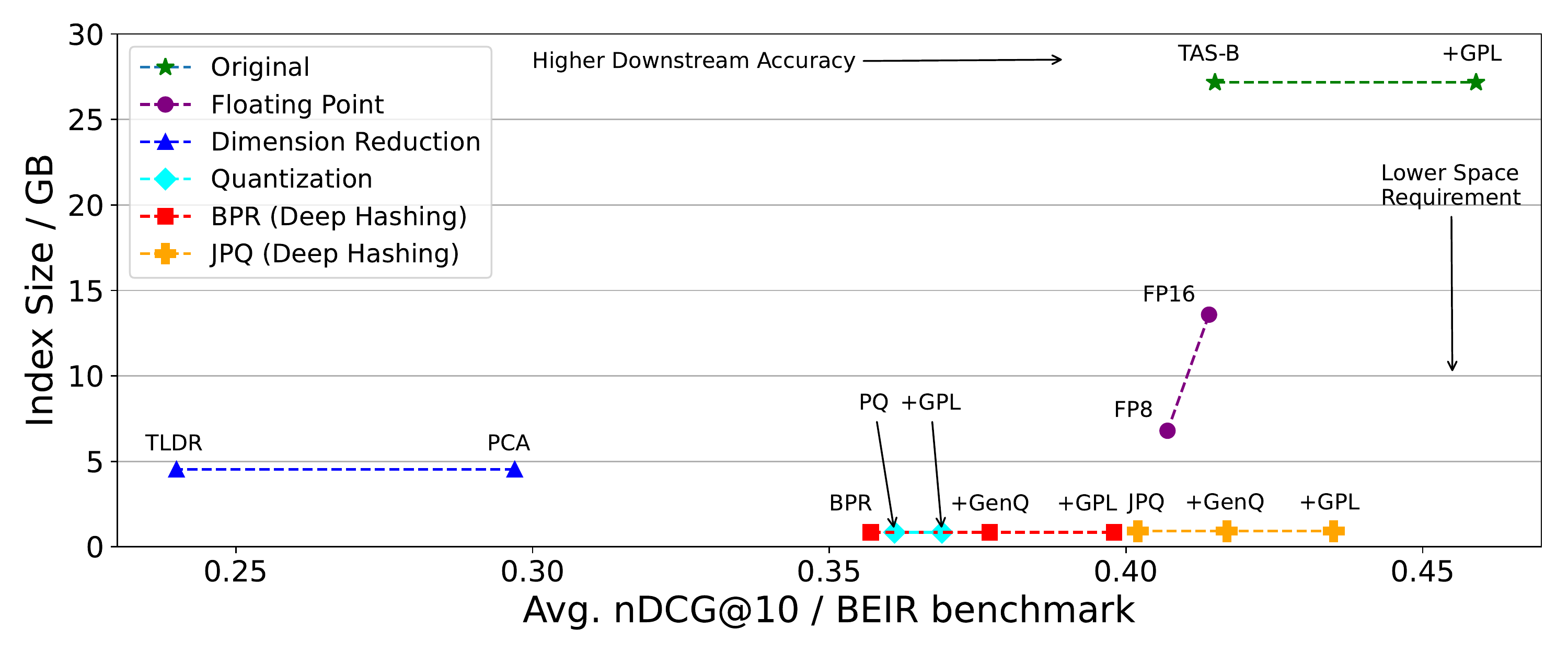}
\captionof{figure}{Summary of index size (estimated for MS MARCO \cite{nguyen2016ms} in GB) versus effectiveness measured as downstream zero-shot retrieval accuracy, i.e. average nDCG@10 on BEIR \cite{thakur2021beir} for all vector compression models evaluated in our work. A higher nDCG@10 and lower index size is desired. Injecting GPL improves BPR and JPQ (zero-shot) by 11.5\% and 8.2\% respectively without any increase in index size.}
\vspace{-8mm}
    \label{fig:overall-results}
\end{center}
\end{figure*}

To resolve this, in our work, we propose an easy yet effective solution of injecting domain adaptation with existing supervised LTH strategies to improve zero-shot generalization without sacrificing efficiency at inference (both memory and retrieval latency).
We focus on the unsupervised setting and experiment with two popular synthetic training data generation-based techniques: GenQ \cite{thakur2021beir} and GPL \cite{wang2021gpl}. 
For both techniques, we use MS MARCO \cite{nguyen2016ms} as the source and keep the rest of the BEIR datasets as the target, as previously experimented in \cite{thakur2021beir, wang2021gpl, ji-xin-modir}. \autoref{fig:overall-results} summarizes the overall results presented in our work.
Our injection technique can improve the naive implementations for both the BPR and JPQ variants of TAS-B by 11.5\% and 8.2\% nDCG@10 on average while maintaining its $32\times$ memory efficiency and $14\times$ and $2\times$ speedup in retrieval latency in comparison to TAS-B dense retriever on BEIR.

\vspace{-3mm}
\section{Background Knowledge}
\vspace{-2mm}
\subsection{TAS-B Dense Retriever} \label{sec:tas-b}
TAS-B \cite{Hofstaetter2021_tasb_dense_retrieval} is a DistilBERT-based \cite{distilbert-2019} siamese dense retrieval model proposed by~\citet{Hofstaetter2021_tasb_dense_retrieval}. A Balanced Topic Aware Sampling (TAS-B) strategy was implemented to form training batches for optimizing retrieval effectiveness in a data-efficient manner.
It was one of the earliest dense retrieval models to successfully exploit knowledge distillation, using dual supervision from two teacher models: a cross-encoder model and ColBERT \cite{10.1145/3397271.3401075}. The TAS-B model has previously been shown to achieve strong generalization performance on BEIR in~\citet{thakur2021beir}, outperforming ANCE \cite{xiong2020approximate} and DPR \cite{karpukhin-etal-2020-dense}. Due to its strong generalization capabilities, we chose TAS-B for our work.

\subsection{Binary Passage Retriever (BPR)} \label{sec:bpr-loss}
BPR, as introduced by \citet{yamada2021efficient}, is a binary retriever trained under supervision in a learning-to-hash setup. BPR builds upon the existing dual-encoder architecture ($f_\theta$; parameterized by $\theta$) and adds a hash layer, which implements a $\mathrm{sign}(\cdot)$ function to convert the dense embedding $f_\theta(p)$ into a binary hash $h(p) = \mathrm{sign}(f_\theta(p))$ where $h(p)\in\{-1,1\}^{dim}$. This architecture helps us save memory as it reduces the size of the dense index by storing lightweight binary hash values in a corpus. During inference, similarity scores are computed in two stages during inference using a retrieve and rerank design. First, BPR retrieves top-$k$ passages for an input query hash $h(q)$ and across all hashed passages $h(p)$ using hamming loss. Next, BPR reranks the top-$k$ passage hashes via dot product with the query embedding \{$\langle e(q), h(p_i) \rangle | i \leq k $\}. 


BPR is trained to represent the binary passage hashes $h(p)$ by optimizing both ranking and contrastive losses. The ranking loss function with the approximated query hash $\tilde{h}({q})$ and passage hash $\tilde{h}({q})$ is defined as: 
\begin{align}
\label{eq:hash-loss}
\begin{split}
\mathcal{L}_{\mathrm{rank}}=\sum_{j=1}^{n-1}\mathrm{max}(0, -(\tilde{h}({q_i})\cdot \tilde{h}({p^+_i})
- \tilde{h}({q_i}) \cdot \tilde{h}({p^-_{i,j}}))+\alpha)
\end{split}
\end{align}
where $\alpha$ is a ranking constant. Similarly, the infoNCE \cite{oord2019representation} loss function $\mathcal{L}_{\mathrm{infoNCE}}$ teaches the model to improve reranking with only passage hashes (second-stage inference). The loss function with the original query embedding $e({q})$ is defined as:
\begin{equation}
\label{eq:rerank-loss}
\mathcal{L}_{\mathrm{infoNCE}}=-\log\frac{e^{e(q_i) \cdot h(p^+_i)}}{e^{e(q_i) \cdot h(p^+_i)} + \sum_{j=1}^{n-1}e^{e(q_i) \cdot h(p^-_{i,j})}}
\end{equation} 
Finally, the model is trained by optimizing for both losses: $\mathcal{L}_{\mathrm{BPR}} = \mathcal{L}_{\mathrm{rank}} + \mathcal{L}_{\mathrm{infoNCE}}$.

\subsection{Joint Optimization of Query Encoder and Product Quantization (JPQ)}\label{sec:jpq}
JPQ, as introduced by \citet{zhan2021jointly}, is a binary retriever trained to jointly optimize the query encoder and PQ index together. In traditional PQ, due to the separation between encoding and compression, the PQ training cannot benefit from the first-stage supervised ranking. JPQ overcomes this challenge and learns to optimize the position of the PQ centroid embeddings, by changing the centroid embeddings to push the relevant document closer to the query embedding and the irrelevant documents further away. At inference, the similarity is computed between the reconstructed quantized passage hash from trained PQ (quantized) centroid embeddings $h(p)$ and the original query embedding $e(q)$. The approximate similarity score is calculated by computing the dot product: $s(q, p) = \langle f_\theta(q), h(p) \rangle$. 

JPQ model is trained to minimize the pairwise loss function $l$ between the query embedding $e(q)$ and reconstructed quantized positive $h(p^+)$ and negative passage $h(p^-)$ using InfoNCE \cite{oord2019representation} loss. The final loss function is given as: \begin{align*}
\mathcal{L}_{\mathrm{JPQ}}=-\sum_{i=0}^{n-1} \sum_{i=0}^{n-1} l(e(q) \cdot h(p^+), e(q) \cdot h(p^-))
\end{align*}\label{eq:jpq}.

\vspace{-5mm}
\section{Domain Adaptation Injection}
As our experimental results will show in Section~\ref{sec:experimental_results}, existing LTH strategies such as BPR are not robust against domain shifts within BEIR. To overcome this, we present an easy and effective approach to inject domain adaptation within existing LTH strategies: BPR \cite{yamada2021efficient} and JPQ \cite{zhan2021jointly}.

\subsection{Injecting GenQ Technique}
The GenQ approach, presented by \citet{thakur2021beir} uses a T5-base model trained on MSMARCO is used to generate top-$k$ synthetic queries $\widehat{q_i}$ for a given passage $p_i$ from each target corpora in BEIR \cite{thakur2021beir} individually. 
For injection within existing learning-to-hash (LTH) systems such as BPR and JPQ, we do not change the underlying training setup. We feed the top-$k$ synthetic query-passage training pairs $(\widehat{q_i}, p_i)$ and use them for finetuning BPR and JPQ to better generalize across out-of-domain retrieval tasks in BEIR.

\subsection{Injecting GPL Technique}

GPL (Generative Pseudo Labeling) presented by \citet{wang2021gpl}, is an extension of the GenQ technique. GPL combines query generation with a pseudo-labeling approach to filter out low-quality generated queries. More formally, for a generated query $\widehat{q}$ from the passage $p^+$, the method first retrieves top-k matching passages $\{p^-_{1}, p^-_{2},..., p^-_{k}\}$ from our corpus using an existing retriever. GPL next randomly samples passages to create the synthetic triplets $(\widehat{q}, p^+, p^-)$. Finally, a cross-encoder $(\mathrm{CE})$ model is used to pseudo-label the similarity score $s^+ = (\widehat{q}, p^+)$ as well as $s^- = (\widehat{q}, p^-)$ and compute the score-margin $\delta = s^+ - s^-$. 

Both BPR and JPQ models train the retriever model using infoNCE loss \cite{oord2019representation} which is found sensitive to synthetic low-quality generated queries generated using GenQ, as shown previously in \cite{wang2021gpl}. Hence, in our work, we improve the existing infoNCE loss and replace it effectively with a MarginMSE loss \cite{hofstätter2021improving}. The first-stage ranking loss function $\mathcal{L}_{\mathrm{rank}}$ for BPR is unchanged. In our work, instead of $\mathcal{L}_{\mathrm{InfoNCE}}$, we will be replacing it in both BPR and JPQ with:
\begin{align*}
\mathcal{L}_{\mathrm{GPL}}=-\sum_{i=0}^{n-1} |\big(e({\widehat{q}_i})\cdot h({p^+_i})-e({\widehat{q}_i})\cdot h({p^-_{i}})\big) - \big(\mathrm{CE}(\widehat{q}_i,p^+_i)-\mathrm{CE}(\widehat{q}_i,p^-_i)\big)|^2,
\end{align*}\label{eq:margin-mse} where $\mathrm{CE}(\widehat{q}_i,p)$ is the pseudo-label or similarity score from a cross-encoder teacher model. Note, we can use both the T5 query generation and cross-encoder model from pre-existing models that have been trained on the source domain like MSMARCO.

\section{Experimental Setup}

\subsection{Datasets and Evaluation} 
For all our experiments, we use the MS MARCO passage ranking dataset \cite{nguyen2016ms} as training data for our BPR and JPQ base implementations. It has 8.8M passages and 532.8K query-passage pairs labeled as relevant in the training set. We use the BEIR benchmark \cite{thakur2021beir} to measure the zero-shot generalization accuracy of our models. The BEIR benchmark contains 18 different zero-shot retrieval evaluation datasets with diverse domains and tasks. Following prior work in \cite{thakur2021beir, wang2021gpl}, we evaluate the zero-shot retrieval accuracy at nDCG@10.

\subsection{Baseline Methods}

In our work, we evaluate four different categories of vector compression baselines: (1) original (without compression), (2) floating point, (3) dimension reduction, and (4) quantization. For all baselines, we use the already finetuned TAS-B \cite{Hofstaetter2021_tasb_dense_retrieval} model as the backbone and evaluate them in a zero-shot retrieval setup across 18 diverse datasets from the BEIR benchmark \cite{thakur2021beir}.

\begin{itemize}
    \item \textbf{Original:} We evaluate two dense retriever baselines: (1) ``zero-shot'' baseline with TAS-B from \citet{thakur2021beir} and (2) ``upper-baseline'' of TAS-B + GPL from  \citet{wang2021gpl}.
    \item \textbf{Floating point}: We reduce the floating-point precision of the zero-shot TAS-B model from the default 32 bits \cite{thakur2021beir} to 16 bits ($\mathrm{fp16}$) and 8 bits ($\mathrm{fp8}$) respectively. 
    \item \textbf{Dimensional reduction:} We fit (1) principle component analysis (PCA) transformation across all embeddings produced by TAS-B for each test dataset in BEIR. We reduce the dimension of the passage embedding from 768 to 128 dimensions. (2) We apply the identical approach for evaluating TLDR by \citet{kalantidis2022tldr}. 
    \item \textbf{Quantization:} We decompose in (1) product quantization (PQ) \cite{jegou2010product} the original $d$ dimensional vector into $k$ centroids, quantized and stored with $n$ bits. 
    We apply product quantization with the zero-shot TAS-B model. (2) We evaluate the TAS-B model with GPL across each dataset in BEIR and apply product quantization (PQ+GPL).
\end{itemize}


\vspace{-3mm}
\subsection{Experimental Setup}

In this section, we describe the implementation details which are necessary for baselines, training learn-to-hash strategies, and domain-adaptation techniques. All our code can be found publicly for reproduction purposes. All our experiments only require a single NVIDIA RTX A6000 GPU resource containing a maximum of 48 GB VRAM. We implement all our models using the HuggingFace’s Transformers package \cite{wolf-etal-2020-transformers}. For all our experiments, we use the TAS-B model\footnote{\href{https://huggingface.co/sentence-transformers/msmarco-distilbert-base-tas-b}{https://huggingface.co/sentence-transformers/msmarco-distilbert-base-tas-b}} as backbone containing a maximum sequence length of $350$ tokens.

\vspace{-3mm}
\paragraph{Baselines.}
We use faiss \cite{johnson2019billion} for implementing all our vector compression baselines, except TLDR. 
For the floating-point baselines, we use the faiss $\mathrm{ScalarQuantizer}$\footnote{\href{https://faiss.ai/cpp_api/struct/structfaiss\_1\_1ScalarQuantizer.html}{https://faiss.ai/cpp\_api/struct/structfaiss\_1\_1ScalarQuantizer.html}} in flat mode for both $\mathrm{fp8}$ and $\mathrm{fp16}$. 
For PCA, we use faiss $\mathrm{PCAMatrix}$\footnote{\href{https://faiss.ai/cpp\_api/struct/structfaiss\_1\_1PCAMatrix.html}{https://faiss.ai/cpp\_api/struct/structfaiss\_1\_1PCAMatrix.html}} in the flat mode and downward project the original dimension of TAS-B model from $768$ into $128$ dimensions. 
For PCA, we normalized both the query and document embeddings and computed cosine similarity with additional whitening. 
We did not find any significant improvement by fitting the MSMARCO dataset with each BEIR test dataset, therefore we trained PCA on each BEIR test corpus individually. For TLDR,\footnote{TLDR repository: \href{https://github.com/naver/tldr}{https://github.com/naver/tldr}} we train one additional multi-layer perceptron (MLP) layer with $2048$ dimensions using 5 nearest neighbors to sample positive pairs and downward project the TAS-B model to $128$ dimensions.
For the PQ implementation, we use faiss $\mathrm{ProductQuantizer}$\footnote{\href{https://faiss.ai/cpp\_api/struct/structfaiss\_1\_1ProductQuantizer.html}{https://faiss.ai/cpp\_api/struct/structfaiss\_1\_1ProductQuantizer.html}} in flat mode. To achieve a compression of 32$\times$, we configure PQ with $k=96$ centroids and $n=8$ bits respectively. We also experimented with OPQ \cite{opq}, however, we found it to underperform naive PQ. Similar to PCA, we did not find any significant improvement by fitting the MSMARCO dataset with each BEIR test dataset, therefore we train PQ on each BEIR test corpus individually.

\vspace{-3mm}
\paragraph{Learning-to-Hash techniques.} For training BPR and JPQ models, we first replicate the original training setup of BPR\footnote{BPR repository: \href{https://github.com/studio-ousia/bpr}{https://github.com/studio-ousia/bpr}} and JPQ\footnote{JPQ repository:\href{https://github.com/jingtaozhan/JPQ}{https://github.com/jingtaozhan/JPQ}} using the TAS-B model as the backbone.
Next, we choose TAS-B model (already finetuned on MS MARCO following \cite{Hofstaetter2021_tasb_dense_retrieval}) as the backbone and further finetune on the MS MARCO triplets\footnote{MS MARCO triplets: \href{https://sbert.net/datasets/msmarco-hard-negatives.jsonl.gz}{https://sbert.net/datasets/msmarco-hard-negatives.jsonl.gz}} publicly available in Sentence Transformers repository to convert the existing TAS-B dense retriever into binary retrievers. As we show in Appendix (\autoref{tab:results-original-vs-tasb}), we find the TAS-B model to outperform original NQ-trained BPR \cite{yamada2021efficient} and STAR-based JPQ \cite{zhan2021jointly} models.\footnote{Our method described is model agnostic, i.e., we can choose any model as the backbone for our method. We intuitively observe that a robust dense retriever model as a backbone in our method provides a higher downstream retrieval accuracy with binary retrievers.} In our experiments, this additional fine-tuning stage is inexpensive and quick and only requires only a maximum of 50K additional steps (4-5 hours in a single NVIDIA RTX A6000 GPU) to convert TAS-B into its binary retriever equivalents, i.e., BPR \cite{yamada2021efficient} and JPQ \cite{zhan2021jointly} as shown in the Appendix (\autoref{fig:training-steps}).

\vspace{-3mm}
\paragraph{Domain adaptation injection.} To generate queries for both GenQ and GPL, we use docT5query \cite{nogueira2019doc2query} model finetuned on MS MARCO and generate at most 3 queries per passage using nucleus sampling with temperature 1.0, $k=25$ and $p=0.95$. 
For BPR+GenQ injection, we follow \cite{thakur2021beir} and finetune the TAS-B model as the backbone for 1 epoch and with a batch size of 75 for each individual BEIR synthetic query-passage dataset. We do not mine hard negatives for training with GenQ, i.e. negatives are randomly sampled within the batch.
For BPR+GPL injection, we follow  \cite{wang2021gpl} and finetune the TAS-B model as the backbone for either 1 epoch or a maximum of 45K training steps with a batch size of $32$. 
To retrieve hard negatives, we use three popular models finetuned on MS MARCO available in the SentenceTransformers repository: (1) \textit{msmarco-distilbert-base-tas-b} (backbone model itself), (2) \textit{msmarco-distilbert-base-v3}\footnote{\href{https://huggingface.co/sentence-transformers/msmarco-distilbert-base-v3}{https://huggingface.co/sentence-transformers/msmarco-distilbert-base-v3}} and (3) \textit{msmarco-MiniLML-6-v3}\footnote{\href{https://huggingface.co/sentence-transformers/msmarco-MiniLM-L-6-v3}{https://huggingface.co/sentence-transformers/msmarco-MiniLM-L-6-v3}}. 
Following \cite{wang2021gpl}, for each model, we retrieve a maximum of 50 negatives and uniformly sample one negative passage and one positive passage for each training query to form one training example. For the cross-encoder teacher, we use the \textit{ms-marco-MiniLM-L6-v2}\footnote{\href{https://huggingface.co/cross-encoder/ms-marco-MiniLM-L-6-v2}{https://huggingface.co/cross-encoder/ms-marco-MiniLM-L-6-v2}} 
model, following \cite{wang2021gpl}. For JPQ+GenQ injection, we follow \cite{zhan2021jointly} and finetune the TAS-B model as the backbone for $2$ epochs with a learning rate of $1e^{-4}$, training batch size of $256$ and top-$200$ PQ-retrieved hard negatives. For JPQ+GPL injection, we follow the same setting as JPQ+GenQ injection and replace the infoNCE loss function with the MarginMSE loss \cite{hofstätter2021improving}. Since labeling all top-$200$ hard negatives is computationally expensive with a cross-encoder at each training step with JPQ, we reduce and only label the top-$25$ hard negatives retrieved using the trainable PQ index at each training step.

\begin{table*}[t!]
    \centering
    \resizebox{\textwidth}{!}{\begin{tabular}{l | c c | c c | c c | c c | c c c | c c c }
        \toprule
        \multicolumn{1}{l}{\textbf{Model ($\rightarrow$)}} &
        \multicolumn{2}{c}{Original} &
        \multicolumn{2}{c}{Floating Point}   &
        \multicolumn{2}{c}{Dim. Reduction}   &
        \multicolumn{2}{c}{Quantization} & 
        \multicolumn{3}{c}{Deep Hashing (BPR)} &
        \multicolumn{3}{c}{Deep Hashing (JPQ)}\\ 
        \cmidrule(lr){1-1}
        \cmidrule(lr){2-3}
        \cmidrule(lr){4-5}
        \cmidrule(lr){6-7}
        \cmidrule(lr){8-9}
        \cmidrule(lr){10-12}
        \cmidrule(lr){13-15}
        \multicolumn{1}{l|}{\textbf{Dataset ($\downarrow$)}} &
        \multicolumn{1}{c}{\textbf{TAS-B}} &
        \multicolumn{1}{c|}{\textbf{+GPL}} &
        \multicolumn{1}{c}{\textbf{\texttt{fp16}}} &
        \multicolumn{1}{c|}{\textbf{\texttt{fp8}}} &
        \multicolumn{1}{c}{\textbf{PCA}} &
        \multicolumn{1}{c|}{\textbf{TLDR}} &
        \multicolumn{1}{c}{\textbf{PQ}} &
        \multicolumn{1}{c|}{\textbf{+GPL}} &
        \multicolumn{1}{c}{\textbf{BPR}} &
        \multicolumn{1}{c}{\textbf{+GenQ}} &
        \multicolumn{1}{c|}{\textbf{+GPL}} &
        \multicolumn{1}{c}{\textbf{JPQ}} &
        \multicolumn{1}{c}{\textbf{+GenQ}} &
        \multicolumn{1}{c}{\textbf{+GPL}} \\ 
         \textbf{Memory}  & {\boldmath$1\times$} & {\boldmath$1\times$} &
{\boldmath$2\times$} & {\boldmath$4\times$} & {\boldmath$6\times$} & {\boldmath$6\times$} & {\boldmath$32\times$} & {\boldmath$32\times$} & {\boldmath$32\times$} & {\boldmath$32\times$} & {\boldmath$32\times$} & {\boldmath$32\times$} & {\boldmath$32\times$} & {\boldmath$32\times$}\\ \midrule
   MS MARCO      & 0.408 & 0.408 & 0.407 & 0.408 & 0.315 & 0.266 & 0.358 & 0.358 & 0.397 & 0.397 & 0.397 & 0.400 & 0.400 & 0.400 \\  \midrule
   TREC-COVID    & 0.481 & 0.700 & 0.481 & \cellcolor{yellow!25}0.484 & 0.324 & 0.287 & 0.426 & \cellcolor{yellow!25}0.552 & 0.400 & \cellcolor{yellow!25}0.535 & \cellcolor{yellow!25}0.593 & \cellcolor{yellow!25}0.601 & \cellcolor{green!15}0.702 & \cellcolor{green!15}0.717 \\
   BioASQ        & 0.383 & 0.442 & 0.359 & 0.357 & 0.153 & 0.097 & 0.253 & 0.311 & 0.285 & 0.353 & \cellcolor{yellow!25}0.388 & 0.326 & 0.362 & \cellcolor{yellow!25}0.384 \\
   NFCorpus      & 0.319 & 0.345 & 0.319 & 0.317 & 0.267 & 0.195 & 0.295 & 0.303 & 0.286 & 0.296 & 0.299 & 0.312 & \cellcolor{yellow!25}0.320 & \cellcolor{yellow!25}0.324 \\
   NQ            & 0.463 & 0.483 & 0.463 & 0.461 & 0.343 & 0.265 & 0.402 & 0.389 & 0.447 & 0.333 & 0.457 & 0.446 & 0.404 & 0.456 \\ 
   HotpotQA      & 0.584 & 0.582 & \cellcolor{yellow!25}0.584 & \cellcolor{yellow!25}0.583 & 0.388 & 0.170 & 0.497 & 0.430 & 0.482 & 0.474 & 0.536 & 0.552 & 0.577 & \cellcolor{green!15}0.593 \\ 
   FiQA-2018     & 0.300 & 0.344 & 0.300 & 0.300 & 0.207 & 0.156 & 0.251 & 0.281 & 0.255 & 0.285 & \cellcolor{yellow!25}0.309 & 0.289 & 0.288 & 0.295 \\
   Signal-1M (RT)& 0.289 & 0.276 & \cellcolor{green!15}0.290 & \cellcolor{yellow!25}0.287 & 0.235 & 0.237 & 0.263 & 0.241 & 0.241 & 0.270 & \cellcolor{yellow!25}0.278 & 0.269 & 0.271 & \cellcolor{yellow!25}0.280 \\
   TREC-NEWS     & 0.377 & 0.421 & 0.377 & 0.377 & 0.241 & 0.134 & 0.306 & 0.331 & 0.338 & 0.367 & 0.268 & 0.366 & 0.363 & \cellcolor{yellow!25}0.381 \\ 
   Robust04      & 0.427 & 0.437 & \cellcolor{yellow!25}0.428 & \cellcolor{yellow!25}0.428 & 0.280 & 0.172 & 0.348 & 0.334 & 0.309 & 0.337 & 0.374 & 0.398 & 0.392 & 0.416 \\
   ArguAna       & 0.429 & 0.557 & 0.429 & 0.321 & 0.333 & 0.374 & 0.326 & 0.348 & 0.316 & 0.348 & 0.354 & 0.389 & \cellcolor{yellow!25}0.453 & \cellcolor{yellow!25}0.443 \\
   T\'ouche-2020 & 0.162 & 0.255 & \cellcolor{yellow!25}0.163 & \cellcolor{yellow!25}0.164 & 0.098 & 0.135 & \cellcolor{yellow!25}0.186 & \cellcolor{yellow!25}0.169 & \cellcolor{yellow!25}0.167 & \cellcolor{yellow!25}0.255 & \cellcolor{green!15}0.259 & \cellcolor{yellow!25}0.176 & \cellcolor{yellow!25}0.179 & \cellcolor{yellow!25}0.220 \\ 
   CQADupStack   & 0.314 & 0.357 & 0.314 & 0.314 & 0.270 & 0.202 & 0.272 & 0.309 & 0.287 & \cellcolor{yellow!25}0.329 & \cellcolor{yellow!25}0.336 & 0.296 & \cellcolor{yellow!25}0.329 & \cellcolor{yellow!25}0.337 \\
   Quora         & 0.835 & 0.836 & 0.835 & 0.835 & 0.773 & 0.829 & 0.825 & 0.813 & \cellcolor{green!15}0.852 & \cellcolor{green!15}0.840 & \cellcolor{green!15}0.853 & 0.830 & \cellcolor{green!15}0.849 & \cellcolor{green!15}0.847 \\
   DBPedia       & 0.384 & 0.384 & 0.384 & 0.384 & 0.304 & 0.235 & 0.351 & 0.322 & 0.335 & 0.304 & 0.348 & 0.370 & 0.352 & \cellcolor{green!15}0.385 \\
   SCIDOCS       & 0.149 & 0.169 & 0.149 & 0.147 & 0.116 & 0.104 & 0.124 & 0.132 & 0.130 & 0.137 & \cellcolor{yellow!25}0.152 & 0.134 & 0.148 & \cellcolor{yellow!25}0.155 \\
   FEVER         & 0.700 & 0.759 & 0.700 & 0.700 & 0.386 & 0.213 & 0.629 & 0.607 & 0.571 & 0.559 & 0.539 & 0.664 & 0.681 & \cellcolor{yellow!25}0.719 \\ 
   Climate-FEVER & 0.228 & 0.235 & 0.228 & 0.227 & 0.105 & 0.103 & 0.192 & 0.167 & 0.177 & 0.201 & 0.204 & 0.194 & 0.192 & 0.226 \\ 
   SciFact       & 0.643 & 0.674 & 0.643 & 0.640 & 0.514 & 0.407 & 0.567 & 0.596 & 0.548 & 0.565 & 0.623 & 0.622 & \cellcolor{yellow!25}0.647 & \cellcolor{yellow!25}0.655 \\ \midrule
   AVERAGE       & 0.415 & 0.459 & 0.414 & 0.407 & 0.297 & 0.240 & 0.361 & 0.369 & 0.357 & 0.377 & 0.398 & 0.402 & \cellcolor{yellow!25}0.417 & \cellcolor{yellow!25}0.435 \\
        \bottomrule
    \end{tabular}}
    \caption{Retrieval accuracy (nDCG@10) by different compression models on BEIR \cite{thakur2021beir} using TAS-B \cite{Hofstaetter2021_tasb_dense_retrieval} as backbone. models starting with (+) denotes addition, for e.g. BPR+GPL or JPQ+GenQ. Any compression model outperforming both the original baselines, i.e., TAS-B (zero-shot) and TAS-B+GPL (upper-bound), is highlighted with \textbf{green}, whereas if a model outperforms any one of them is highlighted with \textbf{yellow}.}
    \label{tab:results}
    \vspace{-0.5cm}
\end{table*}

\vspace{-3mm}
\section{Experimental Results}
\label{sec:experimental_results}

\vspace{-1mm}
\subsection{Zero-Shot Baseline Results on BEIR} 
From \autoref{tab:results}, we observe GPL improves zero-shot TAS-B by a huge margin in nDCG@10 on BEIR (0.415$\rightarrow$0.459) which is consistent with prior work \cite{wang2021gpl}.
\texttt{fp16} and \texttt{fp8} to be strong baselines for upto $2$-$4\times$ light compression within a single point difference nDCG@10 on BEIR. This shows a majority of the relevant information in dense retriever is contained within initial 16 or 8-bit floating decimals. Only ArguAna has a visible drop in nDCG@10 (0.429$\rightarrow$0.321), where we likely suspect useful information is captured in between 8-16 bits of the query vector (which is a long passage in ArguAna). Dimension reduction finds a huge drop in zero-shot retrieval accuracy, and performs worst in terms of Pareto optimality. This indicates that useful information captured within the original 768 dimensions of the TAS-B dense retriever is lost during the downward projection into lower dimensions. For PCA, we observed whitening and cosine similarity have a visible improvement in nDCG@10 on average in BEIR (0.235$\rightarrow$0.297). Contrary to TLDR, we did not observe any improvements with whitening or using cosine similarity. Quantization performs best in terms of Pareto optimality. Contrary to prior work in \cite{opq}, OPQ first conducts a dimensionality reduction and is found to perform worse than PQ in zero-shot retrieval on BEIR. PQ even outperforms zero-shot BPR by a small margin of 0.4 nDCG@10 points. GPL only improves PQ marginally by 0.8 points (0.361$\rightarrow$0.369) nDCG@10 as synthetic queries are added, whereas PQ is trained on domain-specific corpora. 

\vspace{-3mm}
\subsection{Learning-to-Hash Results on BEIR} 
From \autoref{tab:results}, BPR performs competitively in the in-domain retrieval setting on MS MARCO. BPR drops marginally by only 1.1 points nDCG@10 (0.408$\rightarrow$0.397) in comparison with TAS-B dense retriever while gaining $32\times$ memory compression in the tradeoff. Likewise, the JPQ model only shows a drop of 0.8 points nDCG@10 on MS MARCO. 
 This follows that LTH techniques are effective in learning in-domain distributions as shown in prior work \cite{zhan2021jointly, yamada2021efficient}. However, when evaluated zero-shot on BEIR, the performance gap widens. BPR underperforms the TAS-B zero-shot retriever by a substantial margin of 5.8 points nDCG@10. BPR even underperforms the naive quantization PQ baseline by a small margin of 0.4 points nDCG@10. In contrast, we find JPQ as a better robust learning-to-hash (LTH) algorithm. JPQ underperforms the TAS-B dense retriever by 1.3 points and is 4.1 points at nDCG@10 better than BPR. We likely suspect this is due to quantization being more robust to domain shifts, however, an in-depth analysis is required for a better explanation which we keep as future work. 

\vspace{-2mm}
\subsection{Domain Adaptation Injection Results on BEIR} 
We summarize the domain adaptation injection results with GPL and GenQ in \autoref{tab:results}. First, GenQ injection with BPR and JPQ leads to an improvement in specific BEIR datasets which span a single (focused) domain such as FiQA, SciFact, and BioASQ. Datasets in BEIR which cover a broad set of topics (e.g.\ NQ, DBPedia), GenQ performs worse than the BPR, due to the reasons previously outlined in \cite{wang2021gpl, thakur2021beir}. Overall, GenQ is a successful domain adaptation technique and improves both the zero-shot retrieval accuracy of BPR and JPQ by 2.0 and 1.5 points nDCG@10 on BEIR, while maintaining $32\times$ memory efficiency.

Next, we find GPL to be highly effective in domain adaptation. GPL adopts the MarginMSE loss which learns effectively to rank using a cross-encoder teacher. BPR+GPL injection is able to improve zero-shot accuracy on 16/18 datasets, improving the average score of BPR by 11.5\% (0.357$\rightarrow$0.398) nDCG@10 on BEIR. We likely suspect this is due to BPR able to rerank better in the second stage with distilled knowledge from the cross-encoder.
Likewise, we observe similar gains with the JPQ model. GPL injection improves the average score of JPQ by 8.2\% (0.402$\rightarrow$0.435) nDCG@10. GPL injection with JPQ surprisingly even outperforms the original zero-shot TAS-B dense retriever model by 2.0 points nDCG@10 on BEIR. Finally, JPQ+GPL underperforms the upper baseline of TAS-B + GPL by 2.4 points nDCG@10 on BEIR, while gaining a $32\times$ memory efficiency as a Pareto tradeoff.


\begin{table*}[t!]
    \centering
    \caption{Efficiency results reported for all vector compression methodologies measured using query latency (in milliseconds; evaluated on CPU and GPU) and index size (in MB) for two datasets from BEIR \cite{thakur2021beir}: DBPedia \cite{dbpedia-entity} and MS MARCO \cite{nguyen2016ms}. We randomly sample 1M passages from DBPedia for evaluation. We report the mean and standard deviation of retrieval latency. Lower latency and memory are desired.}
    \begin{adjustbox}{max width=\textwidth}
    \begin{threeparttable}
    \begin{tabular}{c | r r r | r r r}
        \toprule
        \multicolumn{1}{l}{} &
        \multicolumn{3}{c}{DBPedia \cite{dbpedia-entity} (1M)} &
        \multicolumn{3}{c}{MS MARCO \cite{nguyen2016ms} (8.8M)} \\
        \cmidrule(lr){2-4}
        \cmidrule(lr){5-7}
        \multicolumn{1}{c}{Model ($\downarrow$)} & \multicolumn{1}{c}{Latency (CPU)} & Latency (GPU) & Index Sizes
        & Latency (CPU) & Latency (GPU) & Index Sizes \\
        \cmidrule(lr){1-1}
        \cmidrule(lr){2-2}
        \cmidrule(lr){3-3}
        \cmidrule(lr){4-4}
        \cmidrule(lr){5-5}
        \cmidrule(lr){6-6}
        \cmidrule(lr){7-7}
        Original & 277.5 $\pm$ 51.81 ms & 7.3 $\pm$ 0.42 ms & 3072.00 MB & 2915.5 + 310.94 ms & 63.1 $\pm$ 0.36 ms & 27162.08 MB \\ 
        PCA & 55.1 $\pm$ 5.61 ms & 4.3 $\pm$ 2.27 ms & 514.76 MB & 447.7 $\pm$ 38.47 ms & 31.3 $\pm$ 1.49 ms & 4529.77 MB \\
        PQ  & 67.6 $\pm$ 5.63 ms & 65.1 $\pm$ 5.07 ms & 96.79 MB & 516.7 $\pm$ 11.71 ms & 524.7 $\pm$ 28.98 ms & 849.60 MB \\
        BPR & 19.7 $\pm$ 2.31 ms &  \multicolumn{1}{c}{n.a.\tnote{a}}  & 96.00 MB & 150.1 $\pm$ 18.78 ms & \multicolumn{1}{c}{n.a.\tnote{a}} &  848.82 MB\\
        JPQ & 170.4 $\pm$ 12.14 ms &  8.0 $\pm$ 1.66 ms & 104.79 MB &  1325.2 $\pm$ 257.64 ms & 75.2 $\pm$ 0.75 ms &  920.34 MB \\
        \bottomrule
    \end{tabular}
    \smallskip\footnotesize
    \begin{tablenotes}
    \item[a] We were unable to evaluate the retrieval latency of the BPR model using the faiss GPU index, as the original authors \cite{yamada2021efficient} optimized the inference code based on numpy. This is also documented here in this \href{https://github.com/studio-ousia/bpr/issues/7}{issue}.
    \end{tablenotes}
    \end{threeparttable}
    \end{adjustbox}
    \label{tab:latencies}
    \vspace{-0.5cm}
\end{table*}

\vspace{-2mm}
\section{Efficiency: Memory and Retrieval Latency}
\vspace{-1mm}
In this section, we compare the efficiency of the vector compression models used in our study. We evaluate our models on two BEIR datasets: (1) 1M sampled passages in DBPedia and (2) MS MARCO. We naively measure the retrieval latency using brute force, i.e. exhaustive search and memory requirements using faiss \cite{johnson2019billion} package. 
We measure the size of the faiss index (in MB) and evaluate the latency on average to retrieve a single query from each dataset (in milliseconds). We conduct experiments on a server with two Intel Xeon Silver 4210 CPUs and a single NVIDIA RTX A6000 GPU.

\vspace{-3mm}
\paragraph{Index sizes.} From \autoref{tab:latencies}, we observe that the TAS-B dense retriever performs well in terms of zero-shot retrieval accuracy, however, significantly increases the index size by several orders of magnitude. On the other hand, models with quantization such as PQ or supervised LTH techniques such as BPR and JPQ and their variants overall achieve a tiny index. All these techniques substantially compress the index and are able to encode MS MARCO \cite{nguyen2016ms} containing 8.8M passages under $1$GB in contrast to TAS-B dense retriever which requires more than $27$ GB.

\vspace{-3mm}
\paragraph{Retrieval latency.} From \autoref{tab:latencies}, we observe that for the CPU setting, overall all compression techniques are much faster compared to brute-force dense retrieval (original) models. PCA provides $5$-$6\times$, PQ provides $4$-$5\times$, BPR provides $14$-$19\times$, and JPQ provides $2\times$ speedup on average in comparison to the TAS-B dense retriever. 
For the GPU setting, we find the brute-force dense retrieval models are significantly faster. TAS-B dense retriever models gain a speedup of $40$-$45\times$ in contrast to CPU inference.
PCA provides a $2\times$ speedup, whereas JPQ gets a comparable retrieval latency in comparison to the original dense retriever. Surprisingly, we find PQ does not improve its latency when evaluated in the GPU setting.

\vspace{-4mm}
\section{Conclusion}
\vspace{-2mm}
Supervised learning-to-hash (LTH) algorithms have been popular and effective across in-domain tasks in retrieval, however, fail to generalize to unseen domains during test time when applied naively. The algorithms are memory efficient but lack downstream accuracy in zero-shot retrieval containing specialized domains for e.g. BEIR benchmark and can severely underperform the zero-shot TAS-B dense retriever. In order to adapt these vector compression algorithms under severe domain shifts, we propose domain adaptation injection in our work. Domain adaptation with GPL improves zero-shot effectiveness of both BPR and JPQ with 11.5\% and 8.2\% nDCG@10, and maintains the $32\times$ memory efficiency and $14\times$ and $2\times$ CPU speedup on BEIR. 

\vspace{-3mm}
\section{Limitations and Future Work}
\vspace{-2mm}
In our work, we find a huge improvement in both efficiency and downstream retrieval accuracy with domain-adaptation injection with BPR \cite{yamada2021efficient} and JPQ \cite{zhan2021jointly}. Our work has a few limitations which we briefly mention and motivate further future work.

\begin{itemize}
\item \textbf{Different compression algorithms.} In our work, we considered JPQ and BPR due to their popularity and effectiveness as shown in our preliminary results. In the future, we can extend our work to newer LTH algorithms, for e.g. RepCONC \cite{10.1145/3488560.3498443} or newer models, for e.g. boosted dense retriever \cite{boosted-dense-retriever}.

\item \textbf{Better backbone models.} We suspect the downstream retrieval accuracy on BEIR can be improved with stronger backbone models than TAS-B, for e.g. Contriever \cite{contriever}. Due to the model-agnostic nature of our method, we can easily extend our work with more robust dense retrievers in the future.

\item \textbf{Requires separate models.} Domain adaptation with BPR and JPQ both require training separate models for each domain with our technique. This can be quite cumbersome in case there are several domains for which the user would need to train multiple models.

\item \textbf{Compute intensive.} Our domain-adaptation technique GPL leads to the highest downstream accuracy but is compute-intensive. For hard negative mining, each model is required to compute embeddings for the whole corpus. Next, the cross-encoder teacher increases model training complexity significantly. In the future, we can explore efficient teachers instead such as ColBERT \cite{10.1145/3397271.3401075} or TILDE \cite{zhuang2021fast}.
\end{itemize}

\vspace{-5mm}
\section{Acknowledgments}
This research was supported in part by the Canada First Research Excellence Fund and the Natural Sciences and Engineering Research Council (NSERC) of Canada. Computational resources were provided by Compute Canada. We would like to thank Kexin Wang for his helpful feedback in earlier drafts of the paper. Additionally, we would like to thank the anonymous reviewers from ReNeuIR 2023 for their helpful feedback and insightful comments.








\bibliography{sample-ceur}




\newpage
\appendix

\begin{figure*}[t]
\centering
\begin{center}
    \includegraphics[trim=0 0 0 0,clip,width=0.4\textwidth]{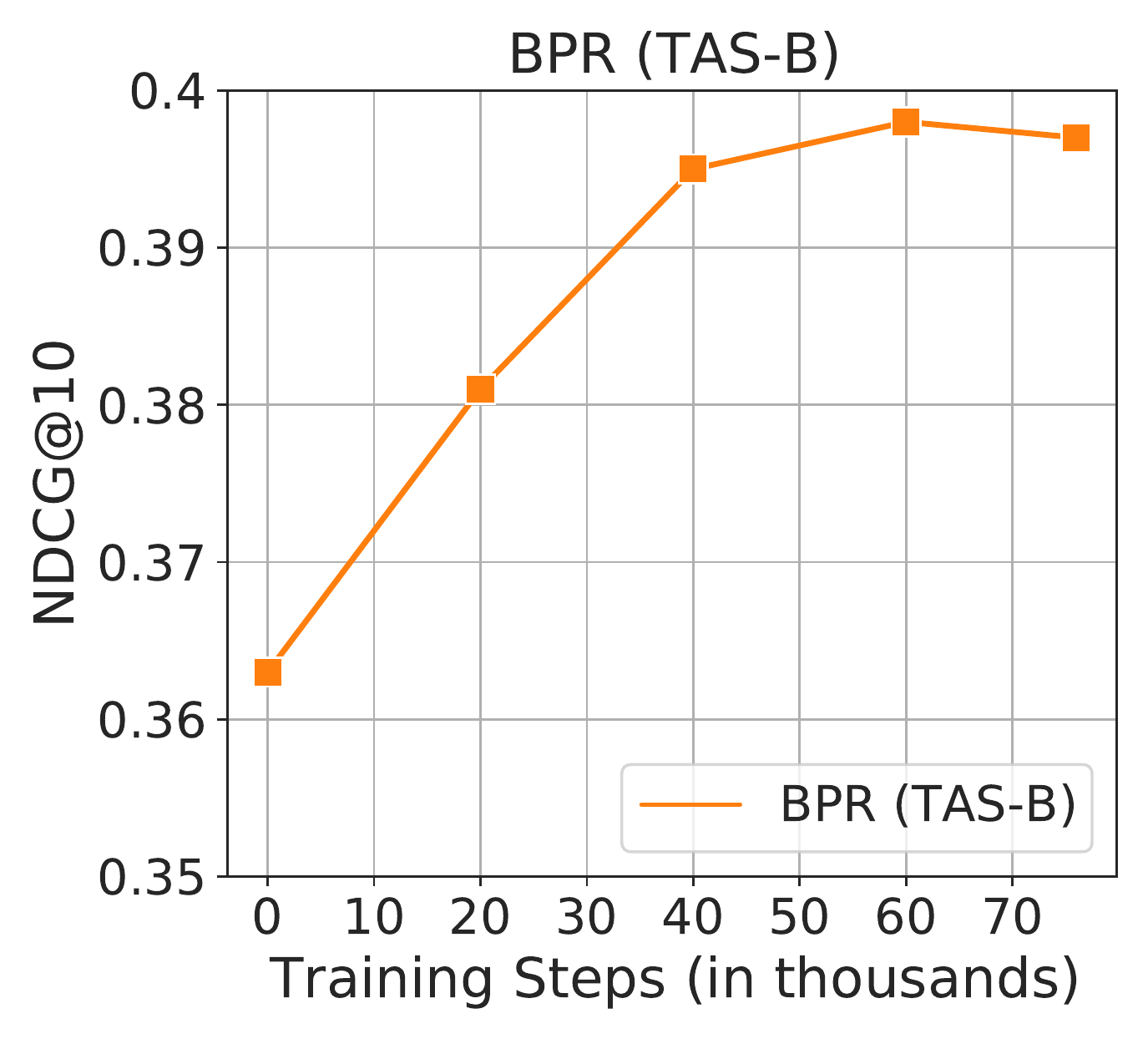}
    \includegraphics[trim=0 0 0 0,clip,width=0.4\textwidth]{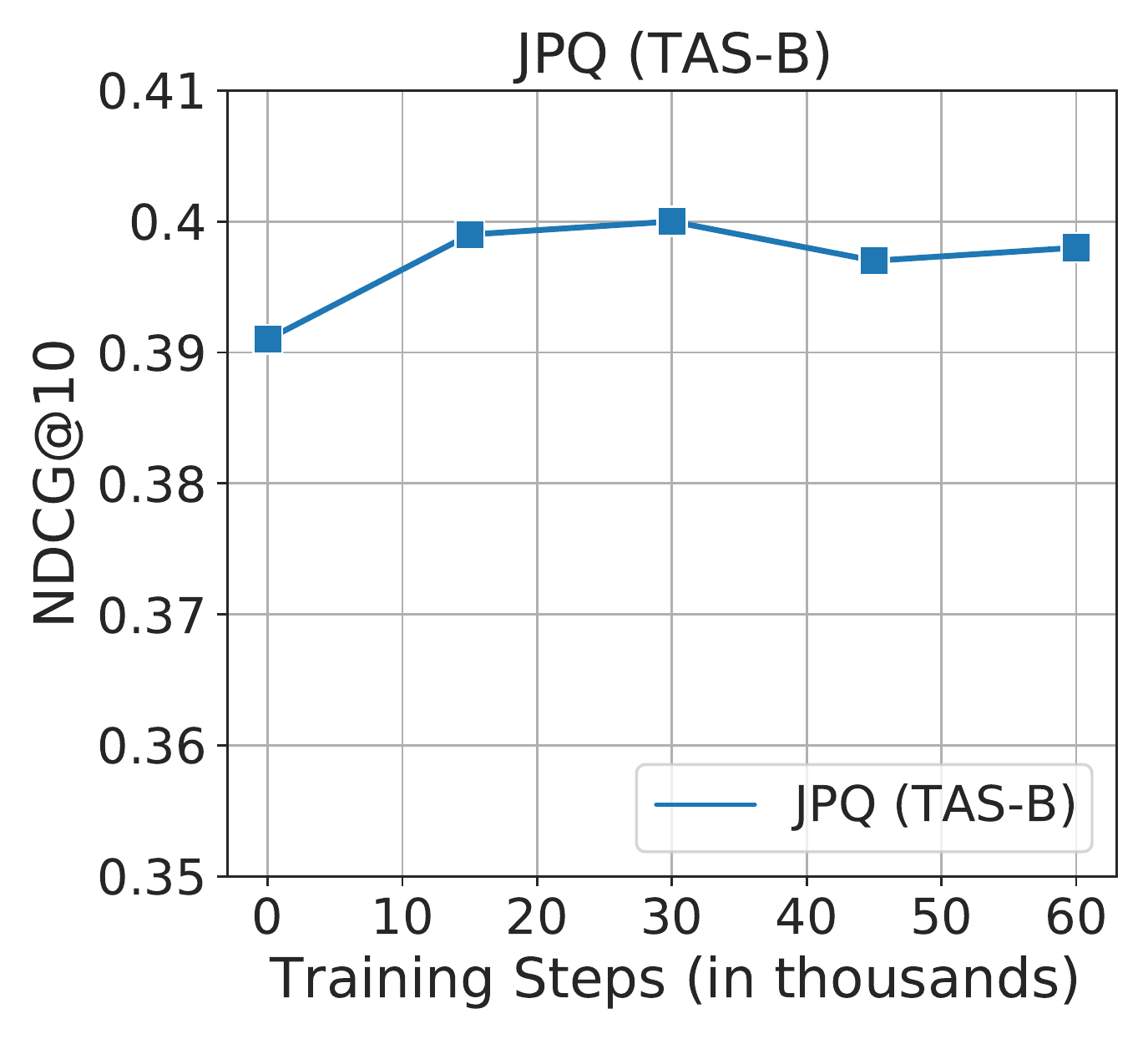}
\captionof{figure}{The number of training steps (in thousands) required to finetune TAS-B \cite{Hofstaetter2021_tasb_dense_retrieval} model for converting into BPR or JPQ. The y-axis denotes the nDCG@10 performance on the MSMARCO DEV \cite{nguyen2016ms} dataset. The training takes approximately 4-5 hours on a single NVIDIA RTX A6000 GPU. The in-domain MS MARCO performance starts to saturate after 30-40K steps.}
    \label{fig:training-steps}
\end{center}
\end{figure*}

\begin{table*}[t!]
    \small
    \centering
    \resizebox{0.8\textwidth}{!}{\begin{tabular}{l | c c | c c}
        \toprule
        \multicolumn{1}{l|}{\textbf{Dataset ($\downarrow$) /Model ($\rightarrow$)}} &
        \multicolumn{1}{c}{\textbf{JPQ (STAR)}} &
        \multicolumn{1}{c|}{\textbf{JPQ (TAS-B)}} &
        \multicolumn{1}{c}{\textbf{BPR (DPR)}} & 
        \multicolumn{1}{c}{\textbf{BPR (TAS-B)}} \\
        \multicolumn{1}{l|}{} &
        \multicolumn{1}{c}{\cite{zhan2021jointly}} &
        \multicolumn{1}{c|}{\textbf{(our work)}} &
        \multicolumn{1}{c}{\cite{yamada2021efficient}} & 
        \multicolumn{1}{c}{\textbf{(our work)}} \\ 
         \textbf{Memory Efficiency} & {\boldmath$32\times$} & {\boldmath$32\times$} & {\boldmath$32\times$} & {\boldmath$32\times$}\\ \midrule
  MSMARCO       & 0.402 & 0.400 & 0.130 & 0.397 \\ \midrule
  TREC-COVID    & \textbf{0.654} & 0.601 & 0.201 & \textbf{0.400} \\
  BioASQ        & 0.306 & \textbf{0.326} & 0.040 & \textbf{0.285} \\
  NFCorpus      & 0.237 & \textbf{0.312} & 0.115 & \textbf{0.286} \\ 
  NQ            & 0.446 & 0.446 & 0.399 & \textbf{0.447} \\ 
  HotpotQA      & 0.456 & \textbf{0.552} & 0.240 & \textbf{0.482} \\ 
  FiQA-2018     & \textbf{0.295} & 0.289 & 0.081 & \textbf{0.255} \\ 
  Signal-1M (RT)& 0.249 & \textbf{0.269} & 0.157 & \textbf{0.241} \\ 
  TREC-NEWS     & \textbf{0.382} & 0.366 & 0.161 & \textbf{0.338} \\ 
  Robust04      & 0.392 & \textbf{0.398} & 0.207 & \textbf{0.309} \\ 
  ArguAna       & \textbf{0.415} & 0.389 & 0.201 & \textbf{0.316} \\  
  T\'ouche-2020 & \textbf{0.240} & 0.176 & 0.073 & \textbf{0.167} \\  
  CQADupStack   & 0.296 & 0.296 & 0.110 & \textbf{0.287} \\
  Quora         & \textbf{0.852} & 0.830 & 0.677 & \textbf{0.852} \\ 
  DBPedia       & 0.281 & \textbf{0.370} & 0.236 & \textbf{0.335} \\ 
  SCIDOCS       & 0.122 & \textbf{0.134} & 0.058 & \textbf{0.130} \\ 
  FEVER         & \textbf{0.669} & 0.664 & 0.325 & \textbf{0.571} \\ 
  Climate-FEVER & \textbf{0.198} & 0.194 & 0.121 & \textbf{0.177} \\ 
  SciFact       & 0.507 & \textbf{0.622} & 0.221 & \textbf{0.548} \\ \midrule
  AVERAGE       & 0.389 & \textbf{0.402} & 0.201 & \textbf{0.357} \\
        \bottomrule
    \end{tabular}}
    \caption{Comparison of TAS-B as the model backbone for JPQ \cite{zhan2021jointly} and BPR \cite{yamada2021efficient} with original model backbones (STAR and DPR). The JPQ (STAR) model was trained on ($M=96$) centroids with STAR \cite{star} (RoBERTa \cite{roberta-2019}) as the backbone. The BPR (DPR) was trained with DPR \cite{karpukhin-etal-2020-dense} as the backbone. All scores denote nDCG@10, \textbf{bold} denotes the best score. The table shows that for both JPQ and BPR, TAS-B as the model backbone outperforms previously used model backbones: STAR and DPR.}
    \label{tab:results-original-vs-tasb}
\end{table*}

\end{document}